\documentclass[12pt]{article}

\def\vp{\varphi}

\newcommand{\Q}{\overrightarrow{Q}}
\renewcommand{\P}{\overrightarrow{P}}
\newcommand{\opi}{\overrightarrow{\pi}}
\newcommand{\ovp}{\overrightarrow{\varphi}}
\newcommand{\Opi}{\hat{\overrightarrow{\pi}}} 
\newcommand{\Ovp}{\hat{\overrightarrow{\varphi}}}

\def\half{\textstyle{\frac{1}{2}}}

\def\H{{\cal H}}

\def\H{{\cal H}}

\def\F{{\cal F}}

\def\ra{\rightarrow}
\def\tint{{\textstyle\int}}

\def\dag{\dagger}

\def\b{\begin{eqnarray*}}  
\def\e{\end{eqnarray*}}    
\def\bn{\begin{eqnarray}}  
\def\en{\end{eqnarray}}   

\def\<{\langle}
\def\>{\rangle}

\def\no{\nonumber}

\def\{{\lbrace}

\def\}{\rbrace}
\begin{document}

\title{ Solving Major Problems \\Using Vector Affine Quantization}          
\author{John R. Klauder\footnote{klauder@ufl.edu} \\
Department of Physics and Department of Mathematics  \\ 
University of Florida,   
Gainesville, FL 32611-8440}
\date{ }
\let\frak\cal
\bibliographystyle{unsrt}

\maketitle

\begin{abstract} 
Affine quantization is a parallel procedure to canonical quantization, which is ideally suited to deal 
with special problems. Vector affine quantization introduces multiple degrees of freedom which find that working together create novel tools suitable to eliminate typical difficulties encountered in more conventional approaches. 
 \end {abstract}

\section{An Introduction to Affine Quantization}
\subsection{Basics of canonical and affine quantization}
The usual variables of classical physics are  $p$ and $q $, and a common Hamiltonian
is $H(p,q)=\half p^2 +V(q)$. If, for some reason $q^2>0$, which is an example of our interest, then we prefer to use the dilation 
$d=pq$ and $q$. For them, the classical Hamiltonian becomes §$H'(d,q) =\half d^2/q^2 +V(q)$. However,
canonical quantization (CQ) requires that
 $-\infty<p\,\&\,q<\infty$, and promotes its proper variables to operators, $p\ra P\,(=P^\dag)$ and $q\ra Q\,(=Q^\dag )$ with
 $[Q,P]=i\hbar1\!\!1$. Affine quantization (AQ) chooses to promote proper variables to
operators, $d=pq\ra D=(P^\dag Q+QP)/2\,(=D^\dag)$, $q\ra Q \neq0$, and $[Q,D]=i\hbar Q$. 
The reason that $Q\neq 0$ is because if $Q=0$ then $D=0$ and $P$ can not help. The fact that $Q\neq0$ leads to either $Q>0$, $Q<0$, or simply both, $Q\neq 0$. Because $Q$ is `incomplete', the operator $P^\dag\neq P$, and so $P^\dag$ was used for $D$ so that $D^\dag=D$ 
However, since $P^\dag Q=PQ$, we find that $D=(PQ+QP)/2$ as well.

  \subsection{An example where affine quantization succeeds}
 As an example of why affine variables are useful consider the well known harmonic operator Hamiltonian,  $\H_{cq}=(P^2 + Q^2)/2=[-\hbar^2(d^2/dx^2)+ x^2]/2$, which successfully  uses CQ. However the half-harmonic oscillator, for which $Q>0$, requires AQ. Since $P^\dag\neq P$, the former Hamiltonian operator fails. 
 Using CQ, we are led to  $ \H_0 = (P P^\dag +Q^2)/2)$  
 AND $\H_1 =(P^\dag P +Q^2)/2$.\footnote{Additional wrong candidate operators include $[(P^\dag+P)^2/4+Q^2]/2$,  $[(2P^\dag-P)(2P-P^\dag)+Q^2]/2$ , $[(P^{\dag 4} /P^2 +P^4/P^{\dag 2})/2+Q^2]/2,$ etc., all having the same classical Hamiltonian!}
 These are TWO different, unequal, solutions, both of which came from the same classical Hamiltonian, $(p^2+q^2)/2$. This is not acceptable quantization! 
 
 Instead, if the problem uses AQ, and we insist that $q>0$, the classical Hamiltonian 
 is $(p^2+q^2)/2=(d^2/q^2+q^2)/2$, and the new, and different, quantum Hamiltonian becomes
  \bn && \H_{aq} = (D\,Q^{-2} D + Q^2)/2 \no \\
     &&\hskip2em = [ -\hbar^2\{[x(d/dx)+(d/dx)x]\,x^{-2}\,[x(d/dx)+(d/dx)x]\}/4+x^2]/2 \no \\
     &&\hskip2em =  [-\hbar^2 (d^2/dx^2) +(3/4)\hbar^2/x^2 +x^2]/2\;, \en 
which has the very same classical Hamiltonian \cite{66}, Sec.~1.5.
While the full harmonic oscillator has equally spaced eigenvalues $\hbar [(0,1,2,..) +1/2))$, the eigenvalues of the half-harmonic oscillator are $2\,\hbar [(0, 1, 2,..) +1]$, which shows a close connection of these two operators \cite{gouba}.

It should be clear that the mathematics of this example would permit us to not only solve the case for $x>0$, but also for $x<0 $, and therefore display two opposing  or different solutions at the same time!
  
This story  is an example of how AQ has solved a problem that CQ cannot solve. The full harmonic oscillator is solved by CQ, and it fails using AQ. Evidently, each of these procedures can solve two different sets of problems.
 
 Let us now introduce vector affine quantization (VAQ) as a lead-up to an illustration of a really difficult problem that VAQ can tackle.

 \section{A First Look at Vector Affine Quantization}
 Individual vectors have multiple similar elements such as  $\P$ and $\Q$ for CQ systems. For AQ
systems there is $D \equiv \half(\P^\dag \cdot  \Q+ \Q\cdot \P) \;(=D^\dag)$ and $\Q\;(=\Q^\dag) \neq 0$.
For the same reason as before, $D=\half(\P\cdot \Q+\Q\cdot \P)$ is also correct. This particular set of basic operators is designed to force $\Q\neq0$. This requirement forces every component 
of $Q_j=0$, for $j=\{1,2,3,...N\}$ for a vector with $N$ components. But. suppose we choose
a different restriction such as $(\Q\cdot\Q -1)\neq 0$, which targets an $N$-dimensional sphere. This requires a different version of D. Since  D terms can be different, let us agree that a d-item pertains to a classical issue, and a D-item pertains to a quantum issue. So far we have only introduced D-items.

We now introduce the general scalar D-item to use that aspect along with {some $\F(\Q)\neq0$, a scalar partner, such as $(\Q\cdot\Q-1)\,((\Q\cdot\Q)^2-4)^2 \neq 0$. Such examples choose to be part of a
D-item leading to 
    \bn &&\hskip-3em {D}=\half[\,|\P| \,(\Q\cdot\Q -1)\,((\Q\cdot\Q)^2 -4)^2 \no \\ &&\hskip4em
     +\,((\Q\cdot\Q)^2- 4)^2\,(\Q\cdot\Q) -1) \,|\P|\,] \;,\en
    where $|\P|^2=\P^\dag\cdot \P$.
    This example offers two spheres that have a common center with one sphere inside the other one.
For this example, there are three distinct regions of interest: (1) outside of the largest sphere, (2) the region between the two spheres, and (3) inside the smallest sphere. As there were different options in Sec.~1.1, we can choose any one  region, or any two regions, or all three regions to  use as contributions to the quantization.

We now turn to vector field theory models that can handle even bizarre vector affine quantization of such examples.

\section{Vector Affine Quantization of Complicated Field Theory Models}
       An absurd vector model field theory, which is ripe for nonrenormalizability using CQ,
       will be found formally soluble using  VAQ. The purpose of studying this model is to show 
       that this `toy model' can be soluble, then surely, less complex problems, that are  more physically relevant, can surely enjoy even simpler solubility.
      
At first we introduce classical vector fields $\opi(x)$ and $\ovp(x)$ and propose an unusual classical Hamiltonian model. To make things more simple let us agree that $\opi(x)^2 \equiv \opi(x)\cdot\opi(x)$
    and $\ovp(x)^2\equiv \ovp(x)\cdot\ovp(x)$. Just to be crazy, let us choose $p$, the power of the interaction term as  $10<p<\infty$, the number of spatial dimensions $s$ as $7<s<\infty$, and the coupling constant $g\geq 0$, all as part of the classical Hamiltonian,
    \bn &&H(\opi,\ovp) =\tint \{ \;\half[ \opi(x)^2+(\nabla{\ovp}(x)^2)  +m^2\,\ovp(x)^2]\no \\
    &&\hskip8em + g\, |(\ovp(x)^2 -1)\,(\ovp(x)^4 -16)|^p\;\}\;d^s\!x \;. \en
While CQ could solve this problem if $g=0$, it would be impossible to solve it when $g>0$. Let us embark on our procedures using VAQ.

First we note an important feature about quantization of all such field theory problems that divergences do not arise because 
 $\ovp(x)=\infty$. Instead, the field $\ovp(x)$ appears in some denominator that briefly vanishes, and with large enough power, i.e., large $p$, integrations -- think path-integration quantization -- can yield divergencies. Our first job is to introduce a d-item, using $|\opi(x)|^2=\opi(x)^2$, and given by
   \bn d(x)=|\opi(x)|\,(\ovp(x)^2-1)\,(\ovp(x)^4-16) \;, \label{w}  \en
   which is a well-designed  d-item. To protect $d(x)$ this process requires that both $(\ovp(x)^2-1)\neq0$ and $(\ovp(x)^4-16)\neq 0$. These two equations define two spheres with the same center,
   and, in the present case, physics tells us to adopt all three sections of open regions. 
   
   Next, we modify the classical Hamiltonian to become
    \bn &&\hskip-3em H'(d,\ovp) =\tint \{ \half[\![ \,d(x)^2\,[(\ovp(x)^2-1)\,(\ovp(x)^4-16)]^{-2} 
    \no \\
    &&+(\nabla{\ovp}(x)^2)  +m^2\,\ovp(x)^2]\!] 
     + g\, |(\ovp(x)^2 -1)\,(\ovp(x)^4 -16)|^p\}\;d^s\!x \;, \en
     which now requires that the interaction term $|(\ovp(x)^2 -1)\,(\ovp(x)^4 -16)|^p \neq 0$.
     So far, this discussion has focussed on the classical story. Now we turn to the quantum story.
     
     The basic operators involved, again with $|\Opi(x)|^2 =\Opi(x)\cdot\Opi(x)$, are 
     \bn &&\hskip-2em \hat{D}(x)=\half \,\{ |\Opi(x)|\,(\Ovp(x)^2-1)(\Ovp(x)^4-16) \no \\
      &&\hskip3em +(\Ovp(x)^4-16)(\Ovp(x)^2-1)\,|\Opi(x)|\,\} \en
   and $\Ovp(x)$ has its several restrictions, namely $\Ovp(x)^2\neq 1$ and $\Ovp(x)^2\neq 4$..
   Note that $\hat{D}(x)$ can be zero thanks to $|\Opi(x)|=0$.
   
   The quantum Hamiltonian is given by
    \bn &&\hskip-0em \H'(\hat{D},\Ovp)= \tint \{\,\half[\![\hat{D}(x)\, [(\Ovp(x)^2-1)(\Ovp(x)^4-16) ]^{-2}\,
    \hat{D}(x) \\. &&\hskip3em 
    +(\nabla{\Ovp}(x)^2)  +m^2\,\Ovp(x)^2]\!] 
     + g\, |(\Ovp(x)^2 -1)\,(\Ovp(x)^4 -16)|^p\}\;d^s\!x \;. \no \en
    First, it follows that 
     \bn  0<\{|(\Ovp(x)^2-1)|\times |(\Ovp(x)^4-16)|\}^{-2}<\infty\;, \en
     and, for $1<p<\infty$, then it also follows that
       \bn 0<\{\,|(\Ovp(x)^2-1)|\times |(\Ovp(x)^4-16)|\,\}^p<\infty \;, \en
       a fact which {\it guarantees} that the quantization, although naturally quite complicated, 
       would lead to a valid result, as it did for the half-harmonic oscillator.
       
       \section{A Homework Problem to \\Sharpen Your Understanding}
       Following the procedures above, please test your understanding by rendering a more 
       natural field theory given by
       \bn H(\pi, \vp) =\tint\{\,\half[\,\opi(x)^2 +(\nabla{\ovp}(x))^2 + m^2\,\ovp(x)^2\,]
        +g\,(\ovp(x)^2)^p\,\}\;d^s\!x\en
        for all $p\geq 8$ and $s\geq6$. This analysis points toward finding  quantum operators
        that are designed to formulate a physically correct quantization. 
        
        Other efforts will be needed to provide 
        desired solutions of the appropriate equations. To give you hope, reread the story of the 
     half-harmonic oscillator in Sec.~1.2, which found its correct formulation and correct solution using AQ.

       \section{Why These Procedures Have Affine Roots}
       Equation (\ref{w}) can be reexamined as
             \bn d(x)=|\opi(x)|\,(\ovp(x)^2-1)\,(\ovp(x)^4-16) \equiv |\opi(x)| \,w(x)\;,  \en
              in which $w(x)=0$ implies that  $(\ovp(x)^2-1)\,(\ovp(x)^4-16) \;=0$.\footnote{It is 
              somewhat extraordinary to imagine that removing a single point in one continuous 
              function it 
              automatically removes two whole spheres in another continuous function.}
           The pair of classical variables, $d(x)=|\Opi(x)| \,w(x)$ and $w(x)$, for each $x$, are 
           simply like $d=pq$ and $q$. 
           
           While classical variables for CQ, namely $p\,\&\,q$,
           are promoted to quantum operators, it is required that they be Cartesian variables, 
           specifically that $d\sigma^2 = A^{-1} dp^2+ A\,dq^2$, in order to get physically
           correct operators \cite{dirac}. On the other 
           hand, to get physically correct quantum operators, the classical variables for AQ, namely 
           $d=pq$ and $q\neq 0$, are required to be from  a 
           constant negative curvature, such as $d\sigma^2=B^{-1} q^2\,dp^2+B\,q^{-2}\, dq^2$,
           where the positive constant $B$ determines the magnitude of the curvature \cite{66}, 
           Sec.~1.4. This relation requires that $q^2>0$, which then applies to
           either  $q>0$,   $q<0$,  or both, $q\neq0$. 
           
           For our principal topic, the classical variables, $d(x)=|\opi(x)|\,w(x)$ and 
           $w(x)\neq 0$ must belong to constant negative 
           curvatures to promote physically correct quantum operators, which arises from the 
           metric 
 \bn d\sigma^2 =\tint \{C(x)^{-1}\; w(x)^2\;d |\opi(x)|^2+ C(x)\;w(x)^{-2}\;d w(x)^2\;\}\;d^s\!x\;,\en
        in which $C(x)>0$ determines the magnitude of the constant negative curvatures for each $x$, 
        and all of $w(x)>0$ and $w(x)<0$ are included in this specific analysis. The fact that our 
        model has $(\nabla{\Ovp}(x))^2$ forces an effective continuity of $w(x)$ along
         spheres.

       \section{Summary}
       
       Canonical quantization tackles many field theory problems with a fixed procedure. 
       On the other hand, this version of vector affine quantization  allows for a 
       carefully-designed operator
       retooling in order to tackle a large variety of Hamiltonians to ensure quantization succeeds as 
       already was the case in properly quantizing the half-harmonic oscillator. Additional
       articles featuring affine quantization can be found in \cite{1, 2, 3}.
       
      Having seen what procedures were used, the reader is welcome to carry out a VAQ of their own 
      favorite, unsolved, quantum field theory problem.

\end{document}